\documentclass[conference, 10pt]{IEEEtran}
\ifCLASSINFOpdf
\else
\fi
\usepackage{amsmath}
\usepackage{amssymb}
\usepackage{epsfig,latexsym}
\usepackage{cite}
\usepackage{graphicx}
\usepackage{color}
\IEEEoverridecommandlockouts

\begin{document}
%

\title{Information Theoretic Bounds Based Channel Quantization Design for Emerging Memories}

\author{\IEEEauthorblockN{Zhen Mei, Kui Cai, and Long Shi}
\IEEEauthorblockA{Singapore University of Technology and Design (SUTD), Singapore, 487372\\
Email: $\left\lbrace \rm{mei}\_zhen, cai\_kui\right\rbrace$ @sutd.edu.sg, slong1007@gmail.com}

\thanks{This work is supported by Singapore Agency of Science and Technology (A*Star) PSF research grant, Singapore Ministry of Education Academic Research Fund Tier 2 MOE2016-T2-2-054, and SUTD SRG grant SRLS15095.}
}


%


\maketitle

\begin{abstract}
Channel output quantization plays a vital role in high-speed emerging memories such as the spin-torque transfer magnetic random access memory (STT-MRAM), where high-precision analog-to-digital converters (ADCs) are not applicable. In this paper, we investigate the design of the 1-bit quantizer which is highly suitable for practical applications. We first propose a quantized channel model for STT-MRAM. We then analyze various information theoretic bounds for the quantized channel, including the channel capacity, cutoff rate, and the Polyanskiy-Poor-Verd\'{u} (PPV) finite-length performance bound. By using these channel measurements as criteria, we design and optimize the 1-bit quantizer numerically for the STT-MRAM channel. Simulation results show that the proposed quantizers significantly outperform the conventional minimum mean-squared error (MMSE) based Lloyd-Max quantizer, and can approach the performance of the 1-bit quantizer optimized by error rate simulations.
\end{abstract}


%
\IEEEpeerreviewmaketitle

\section{Introduction}

As an emerging non-volatile memory (NVM) technology, spin-torque transfer magnetic random access memory (STT-MRAM) features compelling advantages such as nanosecond write/read speed, high scalability, low power consumption and unlimited endurance \cite{hosomi2005novel}. However, the reliability of STT-MRAM is seriously affected by process variations and random thermal fluctuations, resulting in the write errors and read errors \cite{zhang2011stt} that are both asymmetric. In the literature, a (71, 64) Hamming code is adopted to correct a single-bit error in Everspin's 16Mb MRAM \cite{mram}. Recently, more advanced ECCs such as low-density parity-check (LDPC) codes and polar codes are proposed which can achieve a better error rate performance for STT-MRAM \cite{cai2013channel, kui2017cascaded, mei2018magn}.

Channel output quantization is critical for the application of various ECCs for emerging memories such as STT-MRAM. To be compatible with the fast read access time of STT-MRAM, the number of quantization bits must be minimized, since read latency increases exponentially with the sensing precision \cite{campardo2005vlsi}. Quantizer design based solely on simulations is time-consuming and cannot simulate low error rate regions. The conventional minimum mean-squared error (MMSE) based quantization, such as the Lloyd-Max quantizer \cite{lloyd1982least} is far from optimum for the STT-MRAM channel, since it minimizes the MSE of the channel before and after quantization only.
\par In information theory, the channel capacity is a fundamental parameter which represents the highest rate at which information can be reliably transmitted. Another important parameter is the cutoff rate, which governs the union bound on the error probability of maximum likelihood (ML) decoding for a random code \cite{jacobs1965principles}. In the literature, both the channel capacity and cut-off rate have been proposed to guide the design of quantizer, for the additive white Gaussian noise (AWGN) channel \cite{jacobs1965principles, liveris2003quantization} and the flash memory channels \cite{wang2011soft}.

In this work, we first propose a quantized channel model for STT-MRAM. We then analyze various information theoretic bounds of the quantized channel, such as the channel capacity and cutoff rate. We note that the blocklengths of ECCs for emerging memories are usually short ({\it e.g.} a few hundred bits). Recently, \cite{polyanskiy2010channel} proposed the Polyanskiy-Poor-Verd\'{u} (PPV) bound which can accurately predict the maximal achievable channel coding rate with finite blocklengths for discrete memoryless channels (DMCs). Correspondingly, we derive the PPV finite-length performance bound for the quantized STT-MRAM channel. By using these channel measurements as criteria, we design and optimize the 1-bit quantizer numerically for the STT-MRAM channel, which is highly suitable for practical applications. Simulation results demonstrate the effectiveness of proposed quantizer design. To the best of our knowledge, this is the first attempt of applying the PPV bound as a criterion for the channel quantization design.

\section{Quantized Channel Model for STT-MRAM}

\subsection{Preliminaries}
In STT-MRAM, each memory cell consists of a magnetic tunneling junction (MTJ) and an nMOS transistor. The reliability of data in the memory cell is severely affected by MTJ process variations, CMOS process variations and random thermal fluctuations, which leads to the write errors and read errors. It has been widely observed that the write error rate (denoted by $P_1$) for $0\rightarrow 1$ switching is much higher than that for $1\rightarrow 0$ switching (denoted by $P_0$). The read errors consist of the read disturb error and the read decision error. The read disturb error (denoted by $P_r$) is also asymmetric which only occurs in one direction. If the read current is along the write-0 direction, only $1\rightarrow 0$ flipping occurs. On the contrary, if the read current is along the write-1 direction, only $0\rightarrow 1$ flipping can occur. In addition, incorrect sensing of the memory cell resistance state causes the asymmetric read decision error.

Driven by the characteristics of these errors, the work in \cite{kui2017cascaded} proposed a cascaded channel model for STT-MRAM, where the write error and the read disturb error are modeled by a combined binary asymmetric channel (BAC). Considering reading along the write-0 direction, the crossover probabilities of the BAC are given by
\begin{align} \nonumber \label{cross}
p_{0}&=\frac{P_{0}}{2}(1-P_{r}); \quad q_{0}=\left( 1-\frac{P_{0}}{2}\right)+\frac{P_{0}}{2}P_{r} \\
p_{1}&= \frac{P_{1}}{2} + \left( 1-\frac{P_{1}}{2}\right)P_{r}; \quad q_{1}=\left( 1-\frac{P_{1}}{2}\right)\left( 1-P_{r}\right).
\end{align}
For the read decision error, \cite{cai2013channel} showed that both the low and high resistances of the STT-MRAM cell approximately follow Gaussian distributions. Hence, a Gaussian mixture channel (GMC) is used to model the distributions of resistances of the memory cell. The means and variances of the low and high resistances are denoted by $\mu_0$, $\mu_1$, $\sigma_0$, and $\sigma_1$, respectively. The concatenation of the BAC and GMC provides a complete description of the STT-MRAM channel. Following the parameters adopted in \cite{zhang2011stt}, in this work, we consider a 45nm$\times$90nm in plane MTJ under a PTM 45nm technology node, with $\mu_0=1\ k\Omega$, $\mu_1=2\ k\Omega$, and $\sigma_0/\mu_0=\sigma_1/\mu_1$. In all our simulations, we take a write error rate of $P_1=2\times 10^{-4}$, and vary $\sigma_0/\mu_0$ to account for the influence of different quality of the fabrication process on the read decision error \cite{kui2017cascaded}.

\subsection{Quantized Channel Model of STT-MRAM}
\begin{figure}[htbp]
\centering
\includegraphics[height=1.4in,width=2.8in]{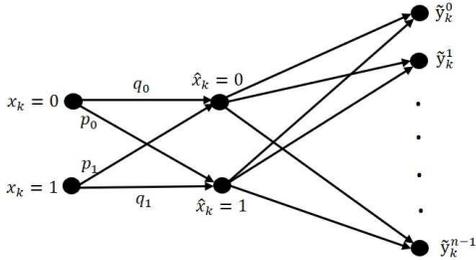}
\caption{Quantized channel model of STT-MRAM.}
\label{channel_q}
\end{figure}
We propose a quantized STT-MRAM channel as shown by Fig. \ref{channel_q}, which is derived based on the cascaded channel model \cite{kui2017cascaded}. This quantized channel consists of a BAC and a binary-input DMC (BI-DMC). The BAC is used to model the combined write errors and read disturb errors and the corresponding crossover probabilities $p_0$, $q_0$, $p_1$, $q_1$ are given by \eqref{cross}. The BI-DMC represents a quantized read decision error channel. In particular, a $q$-bit quantizer maps the channel output signal $y_k$ into $n=2^{q}$ quantized outputs $\tilde{y}_{k}$. Let $a_{0}, a_{1} , \cdots, a_{n}$ be the boundaries of quantization intervals with $a_{0}=-\infty$ and $a_{n}=+\infty$. Define $T_{j}=(a_{j}, a_{j+1})$ as the $j$-th interval, $j=0, 1, \cdots, n-1$. The transition probabilities of the quantized channel are given by
\begin{align} \nonumber \label{Wq0}
W(\tilde{y}_{k}^{j}|x_{k}=0)&= \text{Pr}(\tilde{y}_{k}^{j}|\hat{x}_{k}=0)\text{Pr}(\hat{x}_{k}=0|x_{k}=0)\\ \nonumber
&+\text{Pr}(\tilde{y}_{k}^{j}|\hat{x}_{k}=1)\text{Pr}(\hat{x}_{k}=1|x_{k}=0) \\
&= q_{0}\text{Pr}(\tilde{y}_{k}^{j}|\hat{x}_{k}=0)+p_0\text{Pr}(\tilde{y}_{k}^{j}|\hat{x}_{k}=1)
\end{align}
and
\begin{align} \nonumber \label{Wq1}
W(\tilde{y}_{k}^{j}|x_{k}=1)&= \text{Pr}(\tilde{y}_{k}^{j}|\hat{x}_{k}=0)\text{Pr}(\hat{x}_{k}=0|x_{k}=1) \\ \nonumber
&+\text{Pr}(\tilde{y}_{k}^{j}|\hat{x_{k}}=1)\text{Pr}(\hat{x_{k}}=1|x_{k}=1) \\
&=p_1\text{Pr}(\tilde{y}_{k}^{j}|\hat{x}=0)+q_1\text{Pr}(\tilde{y}_{k}^{j}|\hat{x}_{k}=1),
\end{align}
where $\text{Pr}(\tilde{y}_{k}^{j}|\hat{x}_{k}=i)$ is the transition probability of the read decision channel, given by
\begin{equation}
\text{Pr}(\tilde{y}_{k}^{j}|\hat{x}_{k}=i)= \text{Pr}(y_{k}\in T_{j}|\hat{x}_{k}=i)=\int_{T_{j}}p(y_{k}|\hat{x}_{k}=i)dy_{k},
\end{equation}
with $i=0,1$, $j=0, 1, \cdots, n-1$. Since the read decision error channel is modeled by a GMC, $\text{Pr}(\tilde{y}_{k}^{j}|\hat{x}_{k}=i)$ is calculated as
\begin{equation} \label{pp}
\text{Pr}(\tilde{y}^{j}_{k}|\hat{x}_{k}=i)=Q\left( \frac{a_{j}-\mu_{i}}{\sigma_{i}}  \right)-Q\left( \frac{a_{j+1}-\mu_{i}}{\sigma_{i}}  \right),
\end{equation}
with $i=0,1$, $j=1, 2, \cdots, n-2$. When $j=0$ and $j=n-1$, we have
\begin{equation} \label{pp0}
\text{Pr}(\tilde{y}^{0}_{k}|\hat{x}_{k}=i)=1-Q\left( \frac{a_{1}-\mu_{i}}{\sigma_{i}}  \right),
\end{equation}
and
\begin{equation} \label{ppn1}
\text{Pr}(\tilde{y}^{n-1}_{k}|\hat{x}_{k}=i)=Q\left( \frac{a_{n-1}-\mu_{i}}{\sigma_{i}}  \right).
\end{equation}
By substituting \eqref{pp}, \eqref{pp0} and \eqref{ppn1} into \eqref{Wq0} and \eqref{Wq1}, the transition probabilities of the proposed channel can be obtained. For $1$-bit quantization, the transition probabilities of the read decision error channel are reduced to
\begin{equation} \label{q3}
\text{Pr}(\tilde{y}^{0}_{k}|\hat{x}_{k}=i)=1-Q\left( \frac{a_{1}-\mu_{i}}{\sigma_{i}}  \right)
\end{equation}
and
\begin{equation} \label{q4}
\text{Pr}(\tilde{y}^{1}_{k}|\hat{x}_{k}=i)=Q\left( \frac{a_{1}-\mu_{i}}{\sigma_{i}}  \right).
\end{equation}
 Noted that based on (8) and (9), the transition probabilities of (2) and (3) only depend on $a_1$, the quantization boundary, also known as the decision threshold, of the 1-bit quantizer .

%
%


\section{Information Theoretic Bounds Based Design of The Quantizer}
This section studies various information theoretic bounds of the quantized STT-MRAM channel, based on which we design and optimize the 1-bit quantizer analytically.

\subsection{Capacity-Maximizing Quantizer}

The channel capacity is the maximum code rate that ensures error-free data retrieval in a memory system. In this work, we consider that the input information data follows a uniform distribution with $\text{Pr}(x=0)=\text{Pr}(x=1)=\frac{1}{2}$. The capacity of the quantized channel is then given by
\begin{align} \label{I_q1} \nonumber
C_{q}&=H(Y)-H(Y|X)= -\sum_{j=0}^{n-1}\text{Pr}(\tilde{y}_{k}^{j})\log_{2} \text{Pr}(\tilde{y}_{k}^{j}) +\\
&  \frac{1}{2}\sum_{j=0}^{n-1}\sum_{i=0}^{1}W(\tilde{y}^{j}_{k}|x_{k}=i)\log_{2}W(\tilde{y}^{j}_{k}|x_{k}=i),
\end{align}
where $\text{Pr}(\tilde{y}_{k}^{j})=\frac{1}{2}\left( W(\tilde{y}^{j}_{k}|x_{k}=0) + W(\tilde{y}^{j}_{k}|x_{k}=1) \right).$
After simplifications, $C_{q}$ can be expressed as
\begin{equation} \label{I_q} \nonumber
C_{q}=\frac{1}{2} \sum_{j=0}^{n-1}\sum_{i=0}^{1}W(\tilde{y}^{j}_{k}|x_{k}=i)\log_{2}\frac{W(\tilde{y}^{j}_{k}|x_{k}=i)}{\text{Pr}(\tilde{y}_{k}^{j}) }.
\end{equation}
By maximizing the quantized channel capacity given by \eqref{I_q1}, we can derive the optimal decision threshold $a_1$ for the 1-bit quantizer analytically. In particular, we first compute the derivative of \eqref{I_q1} with respect to $a_1$ as
\begin{align} \label{I_de}
\frac{dC_{q}}{da_{1}}&=H^{'}(Y)-H^{'}(Y|X),
\end{align}
where
\begin{equation}
H^{'}(Y)=-\Psi^{'}\log_{2}\frac{\Psi}{1-\Psi},
\end{equation}
with $\Psi=\frac{q_0+p_1}{2}Q\left( \frac{a_{1}-\mu_{0}}{\sigma_0}\right)+ \frac{p_0+q_1}{2}Q\left( \frac{a_{1}-\mu_{1}}{\sigma_1} \right)$, and
$\Psi^{'}=-\frac{q_0+p_1}{2\sigma_0\sqrt{2\pi}}\exp\left( -\frac{(a_{1}-\mu_0)^2}{2\sigma_{0}^{2}} \right)-\frac{p_0+q_1}{2\sigma_1\sqrt{2\pi}}\exp\left( -\frac{(a_{1}-\mu_1)^2}{2\sigma_{1}^{2}} \right)  $. We also have
\begin{align} \nonumber
H^{'}(Y|X)&= -\frac{1}{2}\sum_{j=0}^{1}\sum_{i=0}^{1} \big( W^{'}(\tilde{y}^{j}_{k}|x_{k}=i)\log_{2}W(\tilde{y}^{j}_{k}|x_{k}=i)\\
& + \frac{1}{\ln2}W^{'}(\tilde{y}^{j}_{k}|x_{k}=i)  \big) ,
\end{align}
where
\begin{eqnarray} \nonumber
W^{'}(\tilde{y}^{0}_{k}|x_{k}=i) =\frac{(1-i)q_{0} + ip_{1} }{\sqrt{2\pi}\sigma_{0}} \exp\left( -\frac{(a_{1}-\mu_{0})^2}{2\sigma_{0}^{2}} \right) \\
+ \frac{(1-i)p_{0} + iq_{1} }{\sqrt{2\pi}\sigma_{1}} \exp\left( -\frac{(a_{1}-\mu_{1})^2}{2\sigma_{1}^{2}} \right),
\end{eqnarray}
and $W^{'}(\tilde{y}^{1}_{k}|x_{k}=i)=-W^{'}(\tilde{y}^{0}_{k}|x_{k}=i)$.
We can then obtain $a_{1}^{*}$ by solving $dC_{q}/da_{1}=0$. In this paper, we use the bisection search to find the optimum threshold $a_{1}^{*}$.

\subsection{Cutoff-Rate-Maximizing Quantizer}
According to \cite{jacobs1965principles}, the cutoff rate of the STT-MRAM quantized channel with equiprobable binary input is given by
\begin{equation} \label{R0_q}
R_{0} \triangleq 1-\log_{2}  \bigg[ 1+ \sum_{j=0}^{n-1}\sqrt{W(\tilde{y}^{j}_{k}|x_{k}=0)W(\tilde{y}^{j}_{k}|x_{k}=1)} \bigg].
\end{equation}
We note that unlike the channel capacity $C_{q}$, $R_{0}$ is achievable and there always exists a finite-length block code with code rate $R$ such that the block error probability (BLEP) $P_{B}$ is less than $2^{-n(R_{0}-R)}$. In this case, the 1-bit quantizer can be optimized by maximizing the cutoff rate. Similar to \eqref{I_de}, we compute the derivative of \eqref{R0_q} with respect to $a_1$.
\par For simplicity, we denote $W^{'}(\tilde{y}^{0}_{k}|x_{k}=0)=\alpha$ and $W^{'}(\tilde{y}^{0}_{k}|x_{k}=1)=\beta$. The derivative of \eqref{R0_q} is given by
\begin{equation} \label{dR}
\frac{dR_{0}}{da_{1}}=-\frac{\frac{\beta\Omega+\alpha\Phi}{2\sqrt{\Omega\Phi}}+ \frac{\beta(\Omega+1)+\alpha(\Phi+1)}{2\sqrt{(\Omega+1)(\Phi+1)}}}{\ln2( \sqrt{(\Omega+1)(\Phi+1)}+\sqrt{\Omega\Phi}+1  ) },
\end{equation}
with $\Omega = p_0\left( Q\left( \frac{\mu_{1}-a_{1}}{\sigma_1} \right)   -1\right) + q_0\left( Q\left( \frac{\mu_{0}-a_{1}}{\sigma_0} \right)   -1\right)$
and
$\Phi = p_1\left( Q\left( \frac{\mu_{0}-a_{1}}{\sigma_0} \right)   -1\right) + q_1\left( Q\left( \frac{\mu_{1}-a_{1}}{\sigma_1} \right)   -1\right)$.
To solve $dR_{0}/da_{1}=0$, the same bisection search method is used.

\subsection{PPV Bound Optimized Quantizer}
Recently, the PPV bound is proposed by \cite{polyanskiy2010channel}, which can closely approximate the maximal achievable channel coding rate at finite blocklength for DMCs. Since the PPV bound is much tighter than the channel capacity and cutoff rate for short-length codes, we propose, for the first time, to design the quantizer by minimizing the BLEP indicated by the PPV bound. According to \cite{polyanskiy2010channel}, given a targeted BLEP $\epsilon$ and for a codeword of length-$N$, the bound of the maximal achievable coding rate can be approximated as
\begin{equation} \label{max_R}
R\approx C_{q}-\sqrt{\frac{V_{q}}{N}}Q^{-1}(\epsilon),
\end{equation}
where the channel dispersion $V_{q}$ is given by
\begin{align} \label{disper_q} \nonumber
V_{q}&=\frac{1}{2} \sum_{j=0}^{n-1}\sum_{i=0}^{1}W(\tilde{y}^{j}_{k}|x_{k}=i) \times  \\ \nonumber
&\left( \log_{2}\frac{W(\tilde{y}^{j}_{k}|x_{k}=i)}{\frac{1}{2}W(\tilde{y}^{j}_{k}|x_{k}=0)+\frac{1}{2}W(\tilde{y}^{j}_{k}|x_{k}=1)}\right) ^2 - C_{q} ^2.
\end{align}
Therefore, for a given $R$, the PPV bound of BLEP is given by
\begin{equation} \label{finite_pe_q}
P_{B}\approx Q\left( \sqrt{\frac{N}{V_{q}}}\left(C_{q}-R \right) \right).
\end{equation}
As it is difficult to derive the close-form derivation of $P_B$ in \eqref{finite_pe_q} with respect to $a_1$, we calculate the derivative numerically to obtain the optimized $a_1$ which minmizes $P_B$.

\subsection{Numerical Results}

\begin{figure}[b]
\centering
\includegraphics[height=2.2in,width=3.3in]{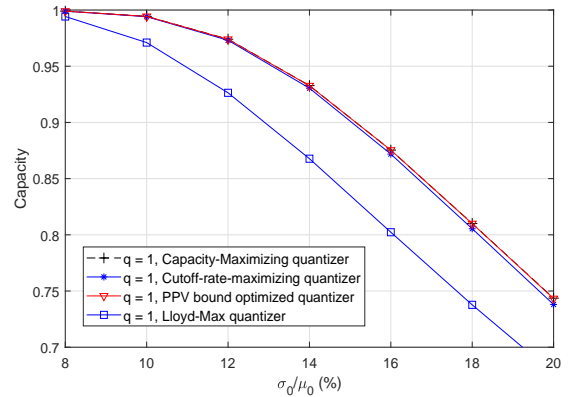}
\caption{The capacity of the STT-MRAM channel with 1-bit quantizer.}
\label{capacity_q}
\end{figure}

\begin{figure}[t]
\centering
\includegraphics[height=2.2in,width=3.3in]{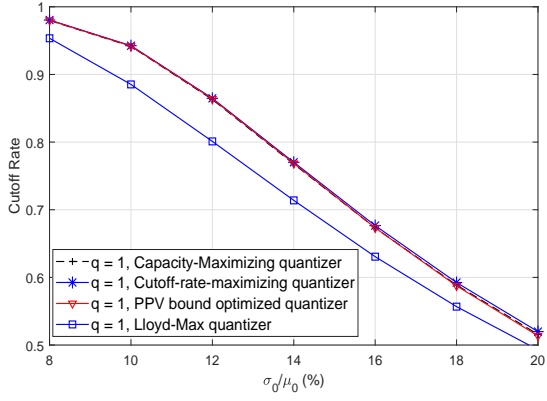}
\caption{The cutoff rate of the STT-MRAM channel with 1-bit quantizer.}
\label{cutoff_q}
\end{figure}

\begin{figure}[t]
\centering
\includegraphics[height=2.2in,width=3.3in]{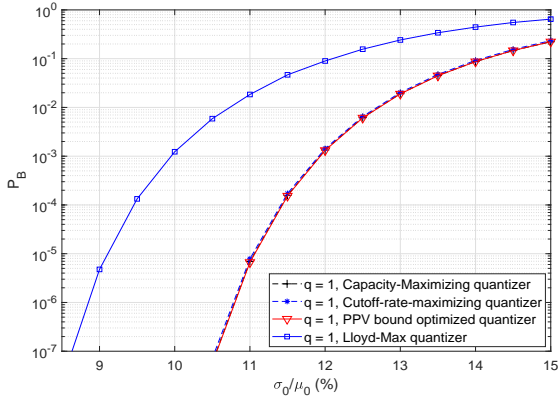}
\caption{The PPV bound of the STT-MRAM channel with 1-bit quantizer.}
\label{finite_q}
\end{figure}

\begin{figure}[t]
\centering
\includegraphics[height=2.2in,width=3.3in]{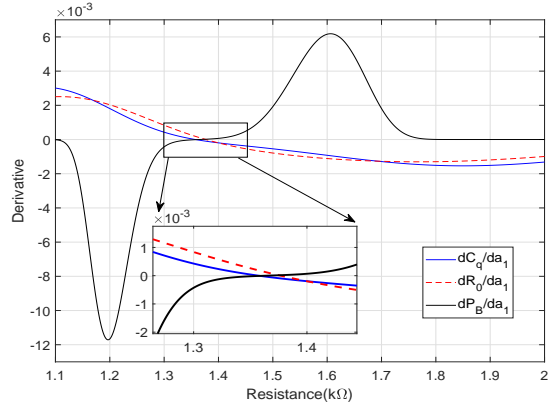}
\caption{The derivative of the capacity, cutoff rate and PPV bound with respect to $a_1$, with $\sigma_0 / \mu_0 =12\%$.}
\label{derivative}
\end{figure}

\begin{figure}[t]
\centering
\includegraphics[height=2.2in,width=3.3in]{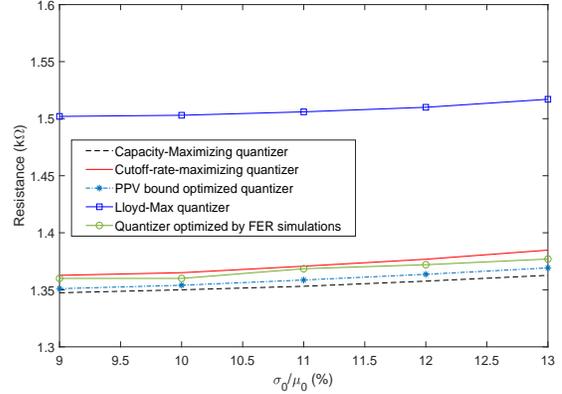}
\caption{The optimum detection thresholds $a_{1}^{*}$ of different 1-bit quantizers, and over different $\sigma_0 / \mu_0$.}
\label{optimal}
\end{figure}
Fig. \ref{capacity_q} depicts the capacities of the STT-MRAM channel with quantizers designed by using the different criteria described above. We observe that the capacity-maximization quantizer achieves a gain of $2\%$ in terms of $\sigma_0 / \mu_0$ over the Lloyd-Max quantizer. The channel capacity with the capacity-maximizing quantizer is slightly higher than that with the cutoff-rate-maximizing quantizer, and is very close to that with the PPV bound optimized quantizer. Fig. \ref{cutoff_q} shows the channel cutoff rates with different quantizers. It is observed that the cutoff-rate-maximizing quantizer outperforms the Lloyd-Max quantizer by $1\%$ in terms of $\sigma_0 / \mu_0$, and the cutoff rates with the other quantizers are very close with each other. The PPV bounds $P_B$ with different quantizers are presented in Fig. \ref{finite_q}. Observe that the PPV bounds with the three quantizers designed based on the information theoretic bounds are similar, which outperform that with the Lloyd-Max quantizer by around $2\%$ in terms of $\sigma_0 / \mu_0$. The derivatives of the capacity, cutoff rate, and the PPV bound of BLEP with respect to $a_1$ are illustrated by Fig. \ref{derivative}. It is observed that optimized detection thresholds using these three criteria are very close with each other, which are away from the midpoint of $\mu_{0}$ and $\mu_{1}$.

\par Fig. \ref{optimal} compares the optimized detection thresholds of different 1-bit quantizers designed above, based on the various information theoretic bounds. The decision threshold of the 1-bit quantizer obtained by simulations by minimizing the frame error rate (FER) of the polar coded channel (see the next section), as well as that of the Lloyd-Max quantizer are also included for comparison. It is observed that the detection thresholds of quantizers designed based on the information theoretic bounds are all close to the optimum detection threshold obtained by simulations, while that of the Lloyd-Max quantizer is far away from the optimum threshold, and it is near the midpoint of $\mu_{0}$ and $\mu_{1}$.


\begin{figure}[t]
\centering
\includegraphics[height=2.5in,width=3.5in]{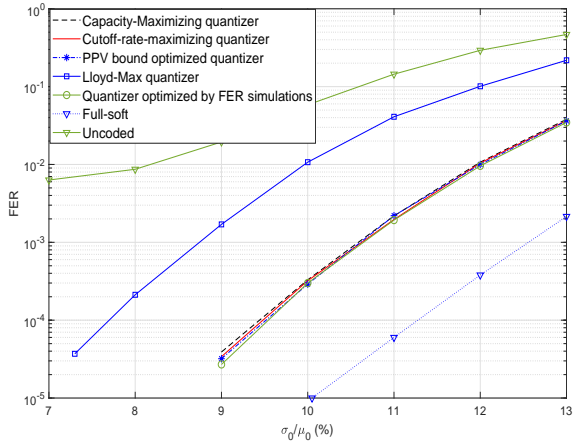}
\caption{FER of the polar code with different 1-bit quantizers.}
\label{FER}
\end{figure}

\begin{figure}[t]
\centering
\includegraphics[height=2.2in,width=3.5in]{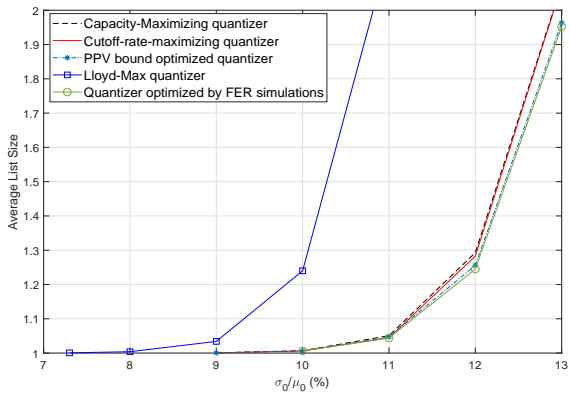}
\caption{The average list size of the AD-SCL decoder of the polar code with different 1-bit quantizers.}
\label{list}
\end{figure}

\section{Simulation Results}
Recently, polar codes have been shown a great potential for STT-MRAM due to its rate-compatible property, flexible decoding algorithms, and excellent error rate performance with short codeword lengths \cite{mei2018magn}. Among various decoding algorithms, the adaptive successive cancelation list decoding (AD-SCL) algorithm achieves the best trade-off between error rate performance and computational complexity. Therefore, in our simulations, we adopt a rate (128, 110) polar code with the AD-SCL decoder. The maximum list size of the AD-SCL decoder is 256 and the number of the cyclic redundancy check (CRC) bits is 4.

\par Fig. \ref{FER} shows the FER performance of the polar code with different 1-bit quantizers. The FERs for the uncoded case, and for the AD-SCL decoder with the full channel soft information are also included as a reference. Observe that the polar code with only 1-bit quantizer achieves a significant coding gain over the uncoded case. The PPV bound optimized quantizer and the cutoff-rate-maximizing quantizer perform slightly better than the capacity-maximizing quantizer, although the performance of all the three quantizers closely approaches that with the quantizer optimized by FER simulations. The corresponding performance are around $2\%$ better the Lloyd-Max quantizer in terms of $\sigma_0 / \mu_0$. It is also observed that the performance of all the 1-bit quantizer (including the quantizer optimized by FER simulations) lags behind the case where the full channel soft information is provided to the polar decoder. This indicates what we have to pay in terms of error rate performance for the simple and fast 1-bit quantizer. It has been verified that this performance gap can be closed if we increase the number of quantization bits to three.

\par Next, we investigate the effect of using different quantizers on the complexity of the AD-SCL decoder, by normalizing it to the size of the simplest successive cancelation (SC) decoder. Fig. \ref{list} shows the average list size of the AD-SCL decoder with different quantizers, and over different resistance spreads $\sigma_0 / \mu_0$. Observe that the average list size of the AD-SCL decoder with different quantizers reduces as $\sigma_0 / \mu_0$ decreases. The average list size with the PPV bound optimized quantizer is slightly smaller than that with the cutoff-rate-maximizing and the capacity-maximizing quantizers, and the complexity of all the three quantizers is similar to that with the quantizer optimized by FER simulations. With these equalizers, and at $\sigma_0 / \mu_0$ of $10\%$ which corresponds to a raw FER of $8\times 10^{-1}$ (see Fig. 7), the average list size of the AD-SCL decoder converges to that of the SC decoder, since only a small number of decoding attempts is required. Again, it is observed that the AD-SCL decoder with the Lloyd-Max quantizer has the highest complexity among all quantizers.

\section{Conclusions}
In this paper, we have considered the design of channel output quantizer with only one quantization bit, which is highly suitable for high-speed emerging memories such as the STT-MRAM. In particular, we have first proposed a quantized channel model for STT-MRAM, based on which we derived various information theoretic bounds, including the channel capacity, cutoff rate, and the PPV finite-length performance bound. By using the derived information theoretic bounds as criteria, we have designed and optimized numerically the 1-bit quantizer for the STT-MRAM channel. Simulation results show that the proposed quantizers perform significantly better than the Lloyd-Max quantizer. The PPV bound optimized quantizer slightly outperforms the capacity-maximizing and cutoff-rate-maximizing quantizers, and achieves a similar performance with the 1-bit quantizer optimized by FER simulations, in terms of both the FER performance and decoding complexity. This demonstrates the high potential of the designed 1-bit quantizer for practical applications.

\bibliographystyle{IEEEtran}
\bibliography{postdoc_refs}

\end{document}